\documentstyle[psfig]{l-aa} %trying to include the figures within.
\begin{document}
\thesaurus{08(03.13.4; 08.16.4; 09.16.1; 11.04.1)}
\title{Improved simulations of the planetary nebula luminosity function}
\author{R.H.~M\'endez \and T.~Soffner}
\offprints{R.H.~M\'endez}
\institute{Institut f\"ur Astronomie und Astrophysik der Universit\"at
           M\"unchen, Scheinerstr. 1, D-81679 Munich, Germany} 
\date{Accepted November 1996}
\maketitle
\begin{abstract}
We have developed a new procedure for the generation of a planetary nebula 
luminosity function (PNLF), improving on previous work (M\'endez et al.
1993). The procedure is based, as before, on an exponential central star
mass distribution, on H-burning post-AGB evolutionary tracks, and on the
{\it avoidance\/} of nebular models for the calculation of nebular fluxes. 
We have added new post-AGB evolutionary tracks and 
introduced the following improvements: (1) the imperfect analytical
representation of post-AGB evolutionary tracks has been replaced by an
interpolation routine giving a better approximation; (2) we have modified
the distribution of the intensities of [O\,{\sc iii}] $\lambda$5007 
relative to H$\beta$, so that it better imitates the observed distribution,
which we have taken from data in the Strasbourg-ESO catalogue of Galactic 
PNs and other sources; (3) we have adjusted the absorbing factor $\mu$
along the white dwarf cooling tracks, so as to reproduce the observed
PNLF of the Magellanic Clouds and M 31 at fainter magnitudes. In this way
we have produced a PNLF which is more consistent with observed PN properties.
We use this randomly generated PNLF to: 
(1) show as convincingly as possible that most PNs in any real population 
must leak stellar H-ionizing photons;
(2) revise our determinations of the parameter $\mu_{\rm max}$;
(3) discuss the shape of the PNLF, hinting at the possible existence of
a distinct feature 2 magnitudes fainter than the brightest PNs;
(4) attempt estimates of maximum final masses from fits to the PNLFs of the
LMC and M 31. There is marginal evidence of a higher maximum final mass in
the LMC, but this should be confirmed in other galaxies with recent star 
formation. A convincing confirmation will require to increase the sample 
sizes and to extend the PNLFs to at least 1.5 mag fainter than the brightest 
PNs.

\end{abstract}

\section{Introduction}

In a previous paper (M\'endez et al. 1993, in what follows MKCJ93) a procedure
was described for the numerical simulation of the bright end of the planetary 
nebula luminosity function (PNLF). The purpose was to model the PNLF in such
a way as to make it possible to study the effects, on the observed
bright end of the PNLF, of the following: (1) sample size; 
(2) time elapsed since last episode of substantial star formation; and 
(3) incomplete nebular absorption of stellar H-ionizing photons. 
This motivation (in particular the desire
to study optical thickness effects) led to the choice of a method 
characterized by the avoidance of nebular models for flux calculations.
In fact, the spectacular images of several nearby PNs, produced 
by the Hubble Space Telescope, have recently added to the current 
uncertainties concerning the transition time between the AGB and the 
moment when the central star becomes hot enough to ionize the nebula
in supporting our feeling that current nebular modeling 
cannot accurately predict how able is the nebula to 
absorb all the H-ionizing photons from the central star, and how
will this ability evolve with time. This explains our
selection of an approach based, as much as possible, on {\it observed\/}
properties of PNs and their central stars.

In the present paper we introduce several improvements in the PNLF 
simulations, in preparation for the moment when better and deeper PNLFs 
will be obtainable with the new 8m-class telescopes now in construction.
The purpose of these improved PNLF simulations is to further develop the 
PNLF as a tool, not only for the accurate measurement of extragalactic 
distances, but also for studies of the initial-to-final mass relation 
and related mass loss processes in luminous galaxies with and without
recent star formation. There has been abundant research on the reliability
of the bright end of the PNLF as a secondary distance indicator (see e.g.
Jacoby 1997), showing that there is excellent agreement between cepheid and 
PNLF distances. This might be interpreted as an indication that the bright 
end of the PNLF is not significantly affected by population 
characteristics. However, it is also possible (and perhaps more
plausible) to argue (e.g. Feldmeier et al. 1997) that 
most PN searches in galaxies with recent star formation have been made 
in such a way as to avoid severe contamination with H\,{\sc ii} regions,
discriminating in this way against the inclusion of the potentially most 
massive central stars, which would tend to be closer to such regions.
The best way of resolving this ambiguity is through PN searches that 
do not discriminate against H\,{\sc ii} regions, like those in
NGC 300 (Soffner et al. 1996). We believe that a reduction of the 
statistical noise through increased sample sizes, as well as an extension
of the PNLF towards fainter magnitudes, should render population effects 
detectable through differences in the shape of the 
PNLF, as described in MKCJ93. Better simulations will allow us to 
predict more confidently what population effects should be 
detectable in practice. In parallel, the 
constraints derived from comparisons of simulated with observed PNLFs
will test how successful we are in modeling the PNLF using an approach 
partly based on random numbers, and will allow us to further refine
the PNLF simulations. In the end, even if population effects
are not detectable, we may be able to understand why.

Sections 2 to 5 describe the improvements we have introduced in our
simulations, and in Sect.\,5 we show that most PNs in any real population 
must leak stellar H-ionizing photons. Sect.\,6 describes a few consistency 
checks that have been made or can be made in principle. In Sect.\,7 we 
discuss the shape of the PNLF, and in Sect.\,8 we describe an attempt to 
determine maximum post-AGB final masses in the Large Magellanic Cloud
(LMC) and in the bulge of M 31.

\section{The improvements}    % This is Section 2.

The basic idea of the procedure for the PNLF simulation is the same as in 
MKCJ93, to which we refer the reader for details. We generate a set of PNs 
with random post-AGB ages and central star masses. 
The ages are given by a uniform random distribution from 0 to 
30\,000 years, counting from the moment when the central star has 
$T_{\rm eff} = $ 25\,000 K. The central star masses are given by an 
exponential random distribution selected to reproduce the observed white
dwarf mass distribution in our Galaxy (see Fig.\,4 in MKCJ93). 
More recent research keeps showing an exponential tail in the white
dwarf mass distribution (see e.g. Bragaglia et al. 1995), and therefore
we are confident that this feature remains valid in cases with more or
less constant star formation. As we will explain later, cases without
recent star formation are simulated by truncating the mass distribution
at a certain maximum final mass.
For each PN in the set we have, then, a pair of random numbers
giving mass and age of the central star. These random numbers are input 
to a routine that gives the corresponding luminosity $L$ and $T_{\rm eff}$
of the central star. Our first improvement is to replace the analytical
representation of post-AGB tracks used in MKCJ93 by an interpolating routine 
that gives a better approximation. Knowing $L$ and $T_{\rm eff}$ we
calculate, using recombination theory, the H$\beta$ luminosity that the
nebula would emit if it were completely optically thick in the H Lyman
continuum. Then we generate a random number, subject to several conditions
(derived from observations of PNs and their central stars, see MKCJ93), 
for the absorbing factor $\mu$, which gives the fraction of 
stellar ionizing luminosity absorbed by the nebula. Using the absorbing
factor we correct the H$\beta$ luminosity. Our second improvement concerns
the generation of suitable absorbing factors for PNs on cooling tracks.
Finally, we produce [O\,{\sc iii}] $\lambda$5007 fluxes from the previously
derived H$\beta$ fluxes. This is done by generating another random number, 
again subject to several conditions, for the intensity of $\lambda$5007 
relative to H$\beta$. The introduction of more stringent conditions for this 
intensity ratio, derived from a larger database, is our third improvement.
In what follows we describe all these changes in more detail. 
We believe that the description will be more easily followed if we 
discuss the $\lambda$5007 intensities before dealing with the absorbing 
factor $\mu$. Stellar masses and luminosities are expressed in solar units.

\section{Improved representation of post-AGB tracks}            % section 3

As in MKCJ93, we base our simulations on the H-burning post-AGB 
tracks of Sch\"onberner (1989) and Bl\"ocker \& Sch\"onberner (1990).
We have added new H-burning tracks recently published by Bl\"ocker (1995).
Since we were not satisfied with the analytical representation used in 
MKCJ93, especially concerning ages and luminosities on the white dwarf 
cooling tracks, we decided to implement some interpolation procedures. In 
addition to the tracks already mentioned, in order to guide the extrapolation
for masses above 0.94 solar masses, we used an additional track for 1.2
solar masses, obtained by slightly modifying the ages in Paczynski's (1971) 
track so as to make them more consistent with the deceleration of the
evolution that takes place along the Bl\"ocker \& Sch\"onberner 0.836 solar 
mass cooling track. 

We used the available evolutionary tracks to construct a look-up table
giving log $T_{\rm eff}$ and log $L$ for 3000 ages between 0 and 30\,000 
years and for 260 masses between 0.55 and 1.2 solar masses. Let us explain 
the construction of this look-up table;
we used different methods for the required calculations in
different regions of the log $L$ - log $T_{\rm eff}$ diagram. 

For temperatures between 25\,000 and 72\,000 K, where the tracks run almost
horizontally, we plotted (1) log (age) and (2) log $L$ as functions of mass 
for a given $T_{\rm eff}$, using information derived from the known tracks; 
fitted log (age) and log $L$ curves as functions of mass; and derived age 
and $L$ for the 260 masses. The procedure was repeated for 40 temperatures 
in this region.

For luminosities below log $L$ = 3.0, which is the region of the white dwarf
cooling tracks, we plotted (1) log (age) and (2) log $T_{\rm eff}$ as 
functions of mass for a given $L$; fitted log (age) and log $T_{\rm eff}$ 
curves; and derived age and $T_{\rm eff}$ for the 260 masses. The procedure 
was repeated for 30 luminosities in this region.

For the remaining region (the knees of the tracks) we produced
fits along 40 straight lines radiating from a point with 
log $T_{\rm eff}$ = 4.86 and log $L$ = 3.0. These lines cross the tracks 
at approximately right angles. In this case it was of course necessary to 
plot log (age), log $L$ and log $T_{\rm eff}$
as functions of mass, and fit curves for the 3 parameters. From these
curves we obtained age, $L$ and $T_{\rm eff}$ for the 260 masses, along
each of the 40 lines we had defined.

Finally, the full look-up table (log $T_{\rm eff}$ and log $L$ for 260
masses and 3000 ages) was completed for the missing ages using interpolation
along each track.

\begin{figure}               %Figure 1
\psfig{figure=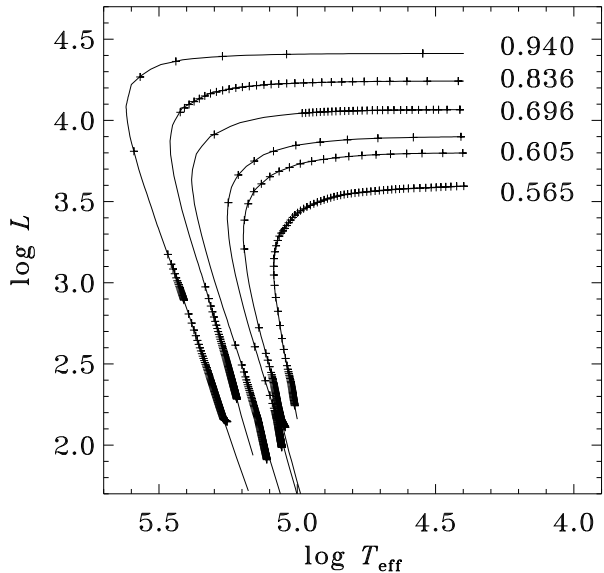,height=8.5cm,angle=90}
\caption[]{
The solid lines are post-AGB evolutionary tracks (H-burning) for 6
different central star masses, taken from Sch\"onberner (1989) and
Bl\"ocker (1995). The unlabeled track is for 0.625 solar masses. 
We used also the track for 0.644 solar masses, but did not plot 
it to avoid overcrowding. The plus signs indicate central star 
luminosities and temperatures calculated, from our look-up table,
for the same 6 masses at 100 post-AGB ages between 1 and 30\,000
years (300-yr intervals). All ages are counted from the moment when
the central star has a temperature of 25\,000 K. For the 3 upper
tracks we have added 30 ages between 1 and 300 yr (10-yr intervals)
to obtain a better coverage of the fast evolution towards higher
temperatures. 
}
\end{figure}

\begin{figure}               %Figure 2
\psfig{figure=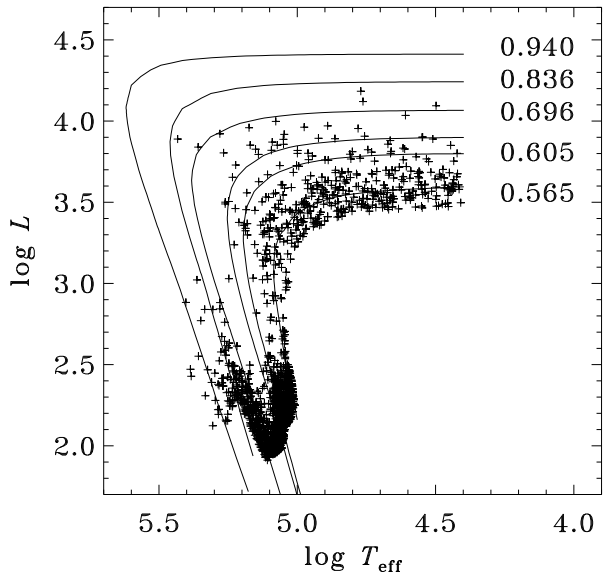,height=8.5cm,angle=90}
\caption[]{
The resulting values of luminosity and temperature for the central stars
of 1500 randomly generated PNs, using the same exponential mass distribution
as in MKCJ93. 
}
\end{figure}

The look-up table was used in the following way: after generating the two
random numbers giving mass and age, the four neighboring values of age 
and mass in the table were
identified, and the values for log $T_{\rm eff}$ and log $L$ were derived
using bilinear interpolation. Fig.\,1 shows examples of evolutionary tracks
generated for several masses, and Fig.\,2 shows the positions of 1500
randomly generated central stars in the log $L$ - log $T_{\rm eff}$
diagram. These two figures can be compared with Figs.\,5 and 6 of MKCJ93.

In addition to a better representation of the post-AGB evolutionary
tracks, the new procedure has the advantage that it is not tied to
a specific set of post-AGB models; it can be applied (i.e. a new look-up
table can be easily generated) for any preferred set of tracks.

We considered, but finally decided against, the inclusion of He-burning 
post-AGB tracks for our simulations. It would be good to allow for a 
certain percentage of such tracks, because the evolution is slower than 
for a H-burning star of the same mass. The basic problem is
that the He-burning tracks are affected by loops and jumps
produced respectively by the late thermal pulse and by the reignition of 
the H-shell (see e.g. Vassiliadis \& Wood 1994,
or Bl\"ocker 1995). It would be necessary to make an enormous amount of
evolutionary calculations in order to accurately reproduce the complexities
of this behavior, which is dependent on the phase, in the thermal pulse 
cycle, at which the star leaves the AGB. No interpolation procedure 
appears to be viable in this case.

The situation would be probably simpler in the particular case of H-deficient
central stars, which represent about 30\% of the well-observed central stars
in our Galaxy (see e.g. M\'endez 1991). Obviously such stars must be 
He-burners, but in this case no H-shell reignition is expected, because
no H is left, and the
tracks may be more easily simulated. The problem is that the evolutionary
status of these objects is not yet fully understood, and no reliable tracks 
are available.

In summary, given the present knowledge we are not likely to gain
much from any attempt to include He-burners in our simulations, but we
remark that this is a potential source of uncertainty in PNLF modeling.

\section{The observed distribution of 
                         $\lambda$5007 relative intensities}   % section 4

One weak point in MKCJ93 was the rather schematic representation of the
distribution of intensities of $\lambda$5007 relative to H$\beta$. We 
have generated a new distribution, by comparison with two observed 
distributions:
one for 118 PNs in the LMC (data taken from Wood et al. 1987; 
Meatheringham et al. 1988; Jacoby et al. 1990;
Meatheringham and Dopita 1991a, 1991b; Vassiliadis et al. 1992) 
and another one for 983 PNs in our Galaxy, taken from the 
Strasbourg-ESO Catalogue of Galactic
PNs (Acker et al. 1992). In about 80 cases, where the Catalogue's
$\lambda$5007 intensity was not given or unreliable, we have taken it
from other sources, listed in the Catalogue.

\begin{figure}                %Figure 3
\psfig{figure=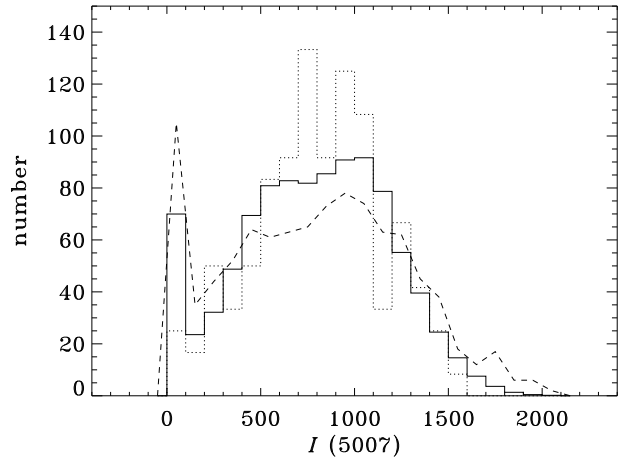,height=6.8cm,angle=90}
\caption[]{
Histograms of the intensity of $\lambda$5007 relative to H$\beta$, on
the scale $I$(H$\beta$) = 100. The dashed line indicates the histogram
for 983 objects in our Galaxy. The other two histograms have been 
normalized to this number. The dotted line is the histogram for 118
LMC objects. The full line is our simulated distribution, generated as 
described in the text.
}
\end{figure}

Fig.\,3 shows the observed distributions compared with our simulated
distribution, which is produced by first generating a Gaussian centered
at an intensity of 1000 (on the usual scale of $I$(H$\beta$) = 100)
with FWHM=300. Then, following the discussion in Section 4 of MKCJ93
about low $\lambda$5007 values (see also Stasinska 1989), we 
decrease the $\lambda$5007 intensity to 50\% of its randomly generated 
value for central stars on heating tracks which have masses smaller than 
0.57 solar masses and $T_{\rm eff} > $ 75\,000 K. In addition, for central 
stars with temperatures below 60\,000 K we do not use the generated random 
number, but use instead Eq.\,(5) in MKCJ93. In this way we minimize the
risk of randomly creating a PN with a $\lambda$5007 intensity which 
is incompatible with the properties of its central star. 

Finally, we must compensate 
for an obvious selection effect: the observed distributions in our Galaxy
and in the LMC are not likely to include PNs with very low-$L$ central stars,
all of which have high temperatures; therefore, before plotting our 
simulated distribution we eliminate all PNs with central stars fainter
than log $L$ = 2.4 (see Fig.\,2). 

We would like to emphasize that the evolutionary 
tracks do not produce enough low-$T_{\rm eff}$ central stars to explain 
the observed number of PNs with low $\lambda$5007 intensities. In order 
to illustrate this point, we can use our simulations. Let us adopt a limit
of 500 for the intensity of $\lambda$5007. According to Eq.\,(5) in MKCJ93,
this limit corresponds to a stellar $T_{\rm eff}$ = 43\,000 K. If we restrict
again our attention to PNs with central stars brighter than log $L$ = 2.4, 
our simulations produce 15\% of these central stars with $T_{\rm eff}$ 
below the limit, while 25\% of the PNs have $\lambda$5007
fainter than 500. This means that about
40\% of the PNs with faint $\lambda$5007 should have hot central stars. 

Since there is insufficient direct information about $T_{\rm eff}$, we 
have tried to test this prediction using the Strasbourg-ESO Catalogue and 
a recent catalogue of $\lambda$4686 line intensities (Tylenda et al. 1994).
We have found that, of 275 PNs in the Strasbourg-ESO Catalogue with an 
intensity of $\lambda$5007 below 500, at least 20\% show the He\,{\sc ii} 
$\lambda$4686 nebular emission, which indicates a stellar temperature 
above 45\,000 K. This percentage is less than the 40\% predicted, 
but on the other hand it is probably a lower limit, because in some 
cases a weak $\lambda$4686 line may be present but not 
detected in the surveys (in the catalogue of Tylenda et al. we find,
among the 275 PNs with weak $\lambda$5007, a further 25\% for which
the upper limit of the $\lambda$4686 intensity is 5 or higher, on the
scale H$\beta=100$). Therefore, although the test cannot be considered 
to be sufficient for any definitive conclusion, given all the uncertainties
involved, we think it indicates that our simulations have produced a
useful first approximation to the observed variety of PN spectra.

As shown in Fig.\,3, we now reproduce the observed distributions quite well.
Notice in particular the Gaussian tail towards high intensities, which was
not correctly reproduced in MKCJ93 (those simulations produced an excess of 
PNs below 1500, and no PN above 1500). 

\section{The absorbing factor $\mu$: reproducing the observed PNLF to fainter 
magnitudes}                                                     % section 5

Assume for a moment that all the PNs are completely optically thick.
We can make a simulation based on this assumption, and compare the 
resulting PNLF with the observed PNLF of the LMC: see Fig.\,4. 
The failure of the completely optically thick assumption is quite evident. 
The simulated PNLF, calculated for a sample size of 1000 objects (which is
Jacoby's (1980) estimate for the total number of PNs in the LMC), reaches 
much brighter magnitudes. In order to force agreement by a horizontal shift
of the observed PNLF, until it fits the simulated PNLF bright end, it would 
be necessary to adopt an implausible distance of 66 kpc for the LMC. An
attempt to fit the observed LMC PNLF at the right distance, by 
{\it vertically\/} shifting the simulated PNLF, would lead to a sample size 
of about 300, implausibly close to the total number of known PNs in the LMC.
This makes it clear that the existence of many optically thin PNs at the
bright end of the PNLF is an essential feature in our simulations. The 
introduction of the absorbing factor $\mu$ in MKCJ93, using information
independently derived from model atmosphere studies of central stars of 
PNs in our Galaxy (M\'endez et al. 1992), immediately led to satisfactory 
fits of the PNLFs of the LMC and M 31. We have not modified the procedure 
described in MKCJ93 for the generation of $\mu$ for PNs with 
high-luminosity central stars (namely, those on heating tracks). 
Here we just remind the reader that many, but not all, of the PNs with
cooler central stars are allowed to have $\mu$=1, and that the procedure 
defines, for central stars on heating tracks and with temperatures above 
40\,000 K, a random distribution of $\mu$ from 0.05 up to a parameter 
$\mu_{\rm max}$. If $\mu_{\rm max}$=1, then a small percentage of the 
bright PNs with hotter central stars can have $\mu$ close to 1. 

\begin{figure}                %Figure 4
\psfig{figure=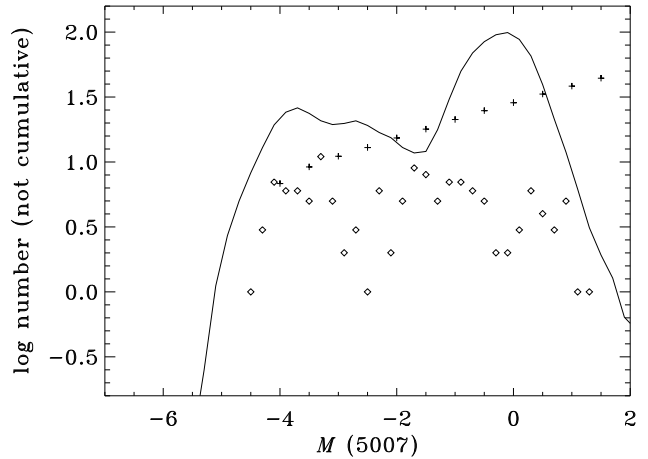,height=6.8cm,angle=90}
\caption[]{
A PNLF simulation for the LMC, assuming that all PNs are completely
optically thick (for all objects $\mu$=1). The diamonds represent the 
observed $\lambda$5007 PNLF of the LMC, from data collected in the 
literature (see text, Sect.\,4). The LMC is assumed to be at a distance of 
50 Mpc, and we adopt an average logarithmic extinction at H$\beta$, $c$=0.19 
(Soffner et al. 1996). The magnitudes fainter than $-3$ are obviously 
affected by severe incompleteness. We have adopted the extension towards 
fainter magnitudes (plus signs) from Fig.\,1 of Ciardullo (1995). The solid 
line is the simulated PNLF, with data binned, like the observed ones, into 
0.2 mag intervals. The sample size is 1000, and the maximum final mass is 
0.7 solar masses (the exponential central star mass distribution is truncated 
at this mass). The choice of this maximum final mass will be explained in 
Sect.\,8. Even with this truncation, the simulated PNLF has too many bright 
PNs and too few of the very faint ones. The discrepancy is solved by forcing 
many PNs to have small $\mu$ values, as shown in Fig.\,5.
}
\end{figure}

Now we want to extend the simulated PNLF towards fainter magnitudes.
This requires some additional information about $\mu$ for PNs with central 
stars on cooling tracks. It is clear from Fig.\,4 that not all PNs on 
cooling tracks can have $\mu=1$, because that assumption produces an 
enormous hump in the PNLF at $M(5007) \sim 0$, and consequently also a 
pronounced deficit of very faint PNs. The required information about $\mu$
cannot be easily obtained from central star studies, because in this region
of the HR diagram the central stars are intrinsically faint.
We have therefore decided to adopt a
simple procedure for the random generation of values of $\mu$ at low
central star luminosities, and adjust it by requiring agreement of the 
simulated PNLF at fainter magnitudes with the statistically complete 
PNLF inferred from PN observations in nearby galaxies (see e.g. Fig.\,1 
in Ciardullo 1995). 

\begin{figure}                %Figure 5
\psfig{figure=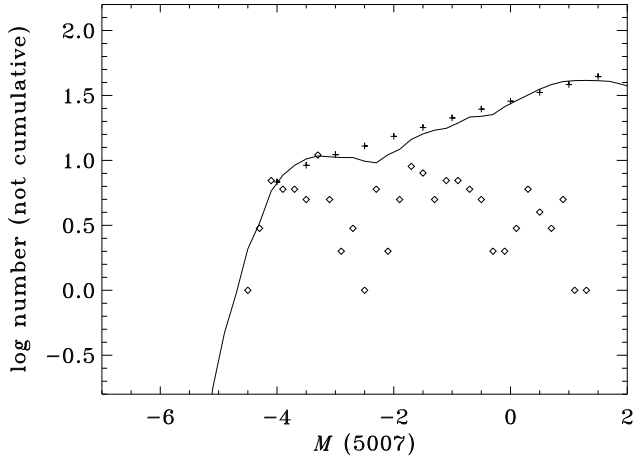,height=6.8cm,angle=90}
\caption[]{
Final simulation of the LMC PNLF.
The same observed LMC PNLF used in Fig.\,4 is now compared with our 
final simulation, which includes many PNs with small $\mu$ values. Most 
of the PNs with high values of $\mu$ have low-temperature central stars.
We have adopted $\mu_{\rm max}$ = 1 (which allows some PNs on heating
tracks and with temperatures above 40\,000 K to be almost
completely optically thick), sample size = 1000, and maximum final 
mass 0.7 solar masses.
}
\end{figure}

The procedure we have implemented is quite simple. We estimate for each
mass how much time it takes for the central star to reach the turnover point
of its evolutionary track, that is to say the beginning of the cooling
track. For all ages larger than this time, the absorbing factor $\mu$ is
set equal to a random number uniformly distributed between 0.1 and 1, 
and this number is multiplied by a factor 
$(1 - ({\rm age(years)} / 30\,000))$. In this way we ensure 
that all absorbing factors tend to 0 as the nebula dissipates. Somewhat 
surprisingly, this simple procedure works very well. Fig.\,5 compares 
the \lq\lq observed'' LMC PNLF with our final simulation, where we have 
set $\mu_{\rm max}$ = 1 for PNs with hot central stars on heating tracks.
The sample size is 1000, in agreement with Jacoby's (1980) estimate, 
and we have selected a maximum final mass equal to 0.7 solar masses 
(see MKCJ93). This choice of maximum final mass will be explained in
Sect.\,8.

It is interesting to note that the factor $(1 - ({\rm age} / 30\,000))$
is necessary to achieve a reasonable representation of the PNLF. If we
suppress this factor, then our simulations give a result very similar
to that shown in Fig.\,4. 

The new value of $\mu_{\rm max}$ for the LMC is higher than determined
in MKCJ93 (0.6). There are two reasons for this change: the new distribution
of $\lambda$5007 intensities, and the adoption of a higher reddening for the
LMC, discussed by Soffner et al. (1996). In next Section we subject our
new simulation to a few consistency checks.

\section{Consistency checks}       %Section 6

In Fig.\,6 we have plotted the H$\beta$ LMC PNLF. In the same way as 
with the $\lambda$5007 PNLF, the simulation provides a good fit at the 
bright end, and severe incompleteness begins about one magnitude fainter. 
This indicates that the simulation is producing suitable values of the 
$\lambda$5007 / H$\beta$ ratio at the bright end of the PNLF. We can 
directly check this by extracting from the master simulation (208\,000 
objects) a random subsample of 1000 objects. In this subsample the 
average $I$(5007) for the 15 brightest PNs is 1181, in good agreement 
with the average intensity of 1213 for the 15 brightest observed PNs 
in the LMC. It is also interesting to know the values of $\mu$ for the
15 brightest PNs in the simulated subsample: they are between 0.38 and 
0.94, with an average 0.74. Thus we see that, although 
$\mu_{\rm max}$=1, the bright end of the PNLF
is {\it not\/} dominated by PNs with $\mu$ values very close to 1.

\begin{figure}                %Figure 6
\psfig{figure=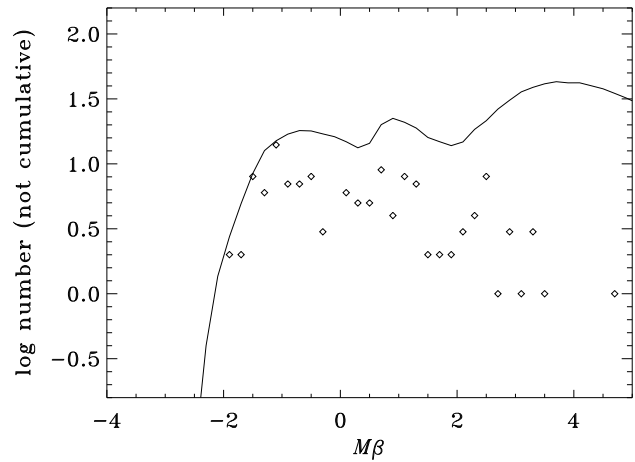,height=6.8cm,angle=90}
\caption[]{
The observed H$\beta$ LMC PNLF, compared with a PNLF built from the 
same simulation used in Fig.\,5: $\mu_{\rm max}$ = 1, 
sample size = 1000, and maximum final mass 0.7 solar masses. There
is a good fit to the bright end of the H$\beta$ PNLF, although here
a somewhat smaller sample size (800) would give a better fit. We find 
severe incompleteness in the observed PNLF for magnitudes fainter than 
$-1$. Thus the situation is very similar to that of Fig.\,5.
}
\end{figure}

This result leads us to rediscuss the interpretation of the observed 
presence of low-excitation PNs among those with the brightest values 
of $M_\beta$ in the LMC (Figure 4a of Dopita et al. 1992). MKCJ93 
concluded that this behavior had to be explained by a low value of
$\mu_{\rm max}$=0.6. But now we see that this behavior can be produced
by chance in cases where $\mu_{\rm max}$ is higher. We have run more 
than 100
simulations for a sample size of 1000 objects and $\mu_{\rm max}$=1, 
and we find that in 30\% of these simulations there are 2 PNs with 
$I(5007) < 500$ among those PNs with the 9 brightest values of 
$M_\beta$ (these numbers, 2 among 9, are what we find in our LMC 
database). For comparison, we have also run more than 100 
simulations for 1000 objects with {\it all\/} values of $\mu$ = 1, 
and in this case less than 7\% of all simulations show 2 
low-excitation PNs among the 9 brightest in H$\beta$.

In this way we see that any argument based on plots of $I(5007)$ as
function of $M_\beta$ makes sense only if applied to a sufficiently
large sample of galaxies. For example, if we were to find that of
10 galaxies the {\it majority\/} show low-excitation PNs among the 
brightest in H$\beta$, then we would be induced to believe that 
something must be inconsistent in our simulations. This is a test
that may become possible in the near future. For the moment we consider
that all the available information is consistent with the following
general rule: many optically thin PNs, but $\mu_{\rm max}$=1, as
implemented in our simulations.

One could argue that the use of $I(5007)$-$M_\beta$ plots to decide 
about $\mu_{\rm max}$ would be further compromised by the possible
existence of low-excitation PNs with hot central stars, as described
in Sect.\,4 above. Although this is true, we should point out that
such objects are generally expected to have very low values of $\mu$, 
and are therefore unlikely to be among the brightest PNs in H$\beta$.
For example, in the case of the LMC we have verified that the 
brightest low-excitation objects (SMP3, SMP5) do not show
He\,{\sc ii} $\lambda$4686 nebular emission. This implies that the 
central star temperatures are indeed low.

\begin{figure}                %Figure 7
\psfig{figure=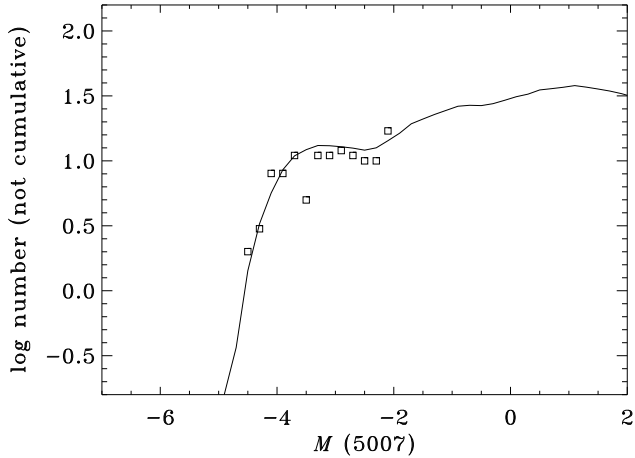,height=6.8cm,angle=90}
\caption[]{
The statistically complete $\lambda$5007 PNLF in M 31 (samples A + B 
of Ciardullo et al. 1989), adopting a distance of 770 kpc, compared 
with a simulated PNLF with $\mu_{\rm max}$ = 1, sample size = 1000, 
and maximum final mass 0.63 solar masses. 
The choice of maximum final 
mass will be explained in Sect.\,8. The change of slope at absolute 
$\lambda$5007 magnitude $-2.3$, predicted by our
simulations, would seem to be reproduced by the data.
}
\end{figure}

\begin{figure}                %Figure 8
\psfig{figure=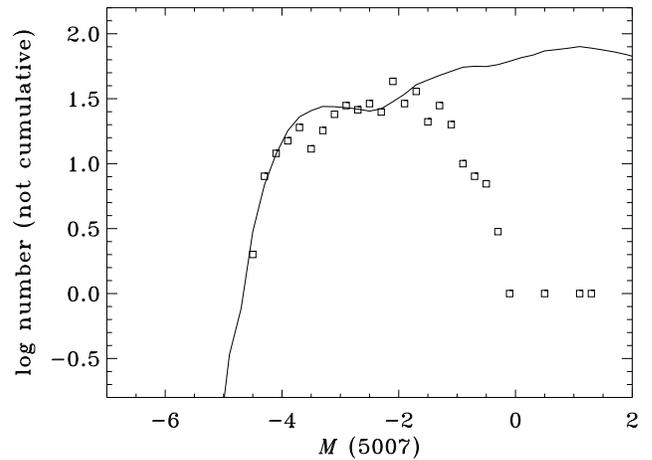,height=6.8cm,angle=90}
\caption[]{
The $\lambda$5007 PNLF of M 31 (all the 429 PNs measured by 
Ciardullo et al. 1989), compared with a simulated PNLF with 
$\mu_{\rm max}$ = 1, sample size = 2100, and maximum final mass 
0.63 solar masses. There is significant incompleteness at mags
fainter than $-1.5$. Notice again the possible change of slope at 
$-2.3$.
}
\end{figure}

\section{On the shape of the PNLF}       %Section 7

Let us compare our simulated PNLF with the formula used by Ciardullo et 
al. (1989, see their Eq.\,(2); their formula behaves like the plus signs 
in our Figs.\,4 and 5). There is an interesting difference: 
while the formula of Ciardullo et al. gives an 
ever increasing PNLF towards fainter magnitudes, our simulation shows a
roughly constant, or even decreasing, PNLF between $M(5007)$ = $-3.5$
and $-2.3$, and only starts increasing again for mags fainter than $-2.3$.
This change of slope can be considered a \lq\lq prediction'' of our 
simulation procedure. The LMC data are not suitable to test this prediction, 
because of the severe incompleteness at the relevant values of $M(5007)$. 
But the M 31 sample remains statistically complete until $M(5007) = -1.5$, 
and indeed Figs.\,7 and 8 would seem to give a hint of support to our 
simulation (see also the combined PNLF shown by Jacoby 1997). 
However, the evidence is insufficient. It would be important to 
verify if the slope change at $ -2.3$ is present in other galaxies 
(this verification will become possible with 8-m telescopes) because 
(1) it would allow to test how reliable is our PNLF generation, 
eventually indicating if further adjustments are needed;
(2) it would give more confidence about how to use the shape of 
the PNLF for distance determinations and for the study of population
characteristics (to be discussed in next section).

The shape of the simulated H$\beta$ PNLF would be
more adequate for shape tests (see Fig.\,6); unfortunately the nebular 
recombination lines (H$\alpha$ would be the natural choice)
are more difficult to measure than $\lambda$5007.

\section{On maximum final masses}                            %Section 8

Fig.\,9 shows our PNLF simulations for different maximum final masses. 
The reason for the different shapes is easy to understand: the more
massive central stars tend to accumulate at the brightest (resp. faintest) 
magnitudes when they are on heating (resp. cooling) tracks. Therefore,
when we eliminate them, the relative percentage of central stars at
intermediate magnitudes (from $-4$ to 0) increases. This dependence of
the PNLF shape on the maximum final mass is what may allow us to learn
about maximum final masses in many different galaxies when suitably
equipped 8-m telescopes become available. It will be a slow trial-and-error
process, because at the same time we will have to test whether or not the
simulated PNLFs produce acceptable and consistent fits. For example,
one critical assumption needed to derive maximum final masses is that we
can use the same value of $\mu_{\rm max}$ for all galaxies. For the
moment, the fact that in the present work we find the same $\mu_{\rm max}$ 
for the LMC and M 31 encourages us to expect this parameter to be valid 
everywhere. But we still have to verify if $\mu_{\rm max}$ = 1 is 
statistically consistent with the morphology of excitation diagrams (plots 
of $I(5007)$ as function of $M_\beta$) made from observations in several 
different galaxies.

Notice that the effect of the maximum final mass is quite different from 
the sample size effect: Fig.\,10 shows, for comparison, simulated PNLFs 
that correspond to several sample sizes. Besides, since a change in 
distance produces a {\it horizontal\/}
displacement of the PNLF, there is no problem for a simultaneous
determination of distance, sample size and maximum final mass, provided 
only that the sample size is large enough to produce small statistical
fluctuations at the bright end of the PNLF. 

\begin{figure}                %Figure 9
\psfig{figure=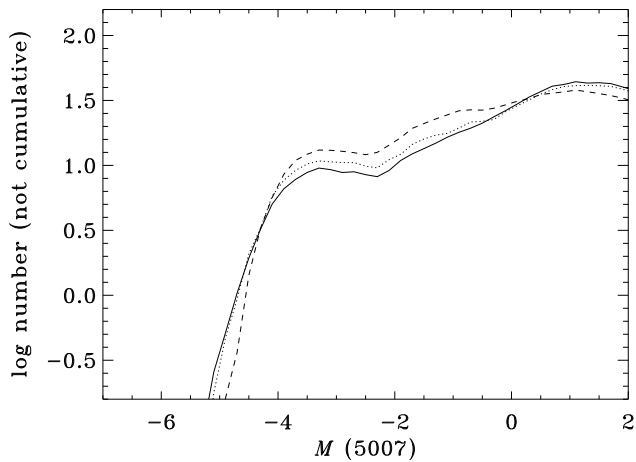,height=6.8cm,angle=90}
\caption[]{
Simulated PNLFs for $\mu_{\rm max}$ = 1, sample size = 1000, and three
maximum final masses: 1.19 (full line), 0.70 (dotted) and 0.63 solar
masses (dashed). In each case the exponential mass distribution has been 
truncated at the limiting mass.
}
\end{figure}

\begin{figure}                %Figure 10
\psfig{figure=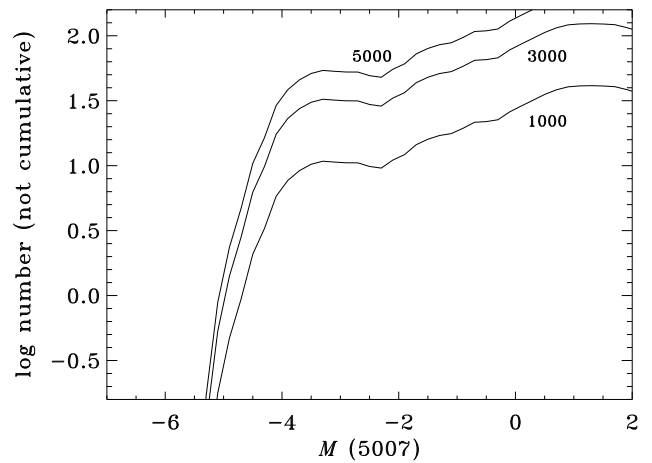,height=6.8cm,angle=90}
\caption[]{
Simulated PNLFs for $\mu_{\rm max}$ = 1, maximum final mass= 0.70 solar
masses, and three sample sizes: 1000, 3000, 5000.
}
\end{figure}

\begin{figure}                %Figure 11
\psfig{figure=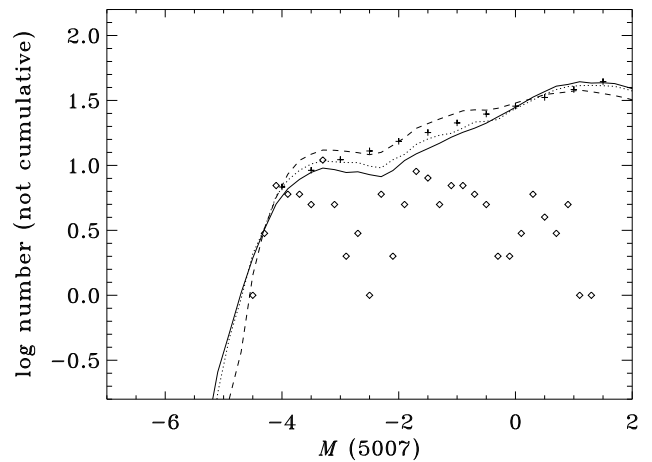,height=6.8cm,angle=90}
\caption[]{
The observed PNLF in the LMC compared with simulations for different 
maximum final masses.
We have overplotted on Fig.\,9 the same LMC PNLF used before. The fit is 
difficult, but it would seem that the best agreement is obtained for
a maximum final mass of 0.70 solar masses (dotted). Notice that the curve 
for 0.63 solar masses (dashed) produces too many PNs at $-3.7$.
}
\end{figure}

\begin{figure}                %Figure 12
\psfig{figure=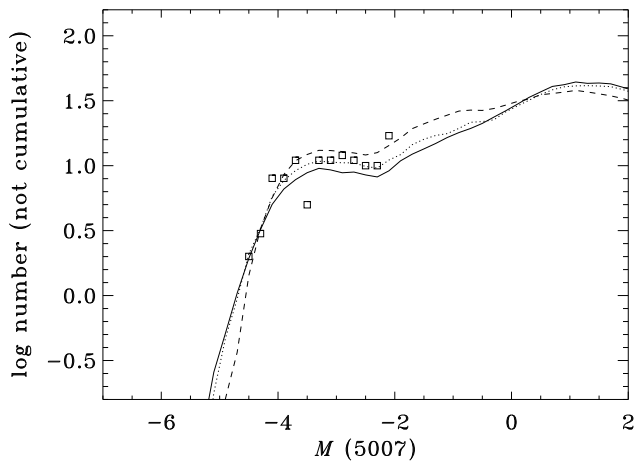,height=6.8cm,angle=90}
\caption[]{
The observed PNLF in M 31 compared with simulations for different maximum 
final masses. We have overplotted on Fig.\,9 the same M 31 PNLF used in 
Fig.\,7. The best agreement is obtained for a maximum final mass of 0.63 
solar masses (dashed).
}
\end{figure}

Using our improved PNLF simulations, we have tried to estimate the maximum
final masses in the LMC and the bulge of M 31. Our preliminary estimates are
0.70 and 0.63 solar masses in the LMC and M 31, respectively; see Figs.\,11
and 12. This would be consistent with the existence of more recent star
formation in the LMC. However, this result is only tentative, due to the 
small number of available bright PNs, and a 
confirmation would be desirable. An improvement of the statistics in the
case of the LMC is rather unlikely, because we cannot expect to find 
many more of the bright PNs; we think it is more promising to
attempt the collection of PN samples in more luminous galaxies 
with recent star formation, and to combine those samples in order to
further increase the sample size.

Although there is quite a lot of information published or in press about PNLFs
in galaxies with recent star formation (see e.g. Jacoby 1997, Feldmeier et 
al. 1997), it is not suitable for the kind of test we would like to make.
Four conditions have to be fulfilled:
(1) there must be abundant evidence of recent star formation. This does
not necessarily imply a restriction to spiral and irregular galaxies: 
consider e.g. the blue bulge of the lenticular galaxy NGC 5102 (McMillan
et al. 1994);
(2) the PN searches must have been made without avoiding the regions of 
recent star formation (for example, we cannot use PNs found in the 
halos of edge-on spiral galaxies like NGC 891 (Ciardullo et al. 1991) and 
NGC 4565 (Jacoby et al. 1996), or in the bulges of M 31 and M 81); 
(3) a statistically complete sample must have been established, extending 
at least 1.5 mag fainter than the bright end of the PNLF, in order to reach
the section of the PNLF where the shape effects are predicted to be most
easily detectable;
(4) to use PNLF distances for this kind of study would be equivalent to 
running in circles. A cepheid distance (or any other universally accepted 
method of distance determination) must be available, in order to eliminate 
any effects derived from the assumption that the PNLF is universal, which 
is used in the application of the maximum likelihood method for PNLF 
distance determinations. 

It turns out that, of the more than 30 galaxies where many PNs have been
found (Jacoby 1997 gives the most up-to-date summary) none satisfies 
simultaneously the 4 conditions. Of course this does not affect in any 
significant way the conclusion that cepheid distances and PNLF distances 
are in excellent agreement; we simply remark that the available samples 
do not allow us to properly study population effects. So far, the galaxy 
that comes closer to fulfill the 4 conditions is NGC 300 (Soffner et al. 
1996), but the sample size must be substantially increased before NGC 300 
can provide a convincing test.

\section{Conclusion}                            % Section 9

One positive aspect of these PNLF simulations is that they make some 
testable predictions, like the possible change of slope at $M(5007)=-2.3$, 
or the relation between the $\lambda$5007 and H$\beta$ (or H$\alpha$) PNLFs, 
or the morphology of the plots of $I(5007)$ as function of $M\beta$ or
$M\alpha$. Several 8m-class telescopes will soon become available, probably 
producing explosive progress in the field of extragalactic PNs. Since there 
will be inevitable efforts to detect many of these objects and obtain their 
spectra, given their usefulness for distance, kinematic and abundance 
studies, we can be assured that the predictions will be testable in the 
near future, leading perhaps to further improvements of the simulated 
PNLFs. Given deeper PN searches and larger sample sizes, we will then have 
better chances to decide if there are indeed detectable population effects 
in the shape of the PNLF, or, if there are not, to understand what are the 
astrophysical reasons for their absence. In any case, through such work the 
PNLF may become a more reliable tool for even more accurate extragalactic
distance determinations.

\section{Acknowledgements}

This work has been supported by the Deutsche Forschungsgemeinschaft through
Grant SFB (Sonderforschungsbereich) 375. We are grateful to J.J. Feldmeier,
R. Ciardullo and G.H. Jacoby for showing us data before publication.

\end{document}